\documentclass[preprint,12pt,authoryear]{elsarticle}

\usepackage{amssymb}
\usepackage{color} 
\usepackage{graphicx}

\journal{New Astronomy}

\begin{document}

\begin{frontmatter}



\title{Modeling the response of a standard accretion disc to stochastic viscous fluctuations}

\author[label1]{Naveel Ahmad}\ead{anaveel@gmail.com}\author[label2]{Ranjeev Misra}\author[label1,label2]{Naseer Iqbal}\author[label1]{Bari Maqbool}\author[label1]{Mubashir Hamid}
\address[label1]{Department Of Physics, University of Kashmir, Srinagar-190006, India}
\address[label2]{Inter-University Center for Astronomy and Astrophysics, Post Bag 4, Ganeshkhind, Pune-411007, India}

\par
\begin{abstract}
The observed variability of X-ray binaries over a wide range of time-scales can be understood in the framework of a stochastic propagation model, where viscous fluctuations at different radii induce accretion rate variability that propagate inwards to the X-ray producing region. The scenario successfully explains the power spectra, the linear rms-flux relation as well as the time-lag between different energy photons. The predictions of this model have been obtained using approximate analytical solutions or empirically motivated models which take into account the effect of these propagating variability on the radiative process of complex accretion flows. Here, we study the variation of the accretion rate due to such viscous fluctuations using a hydro-dynamical code for the standard geometrically thin, gas pressure dominated $\alpha$-disc with a zero torque boundary condition. Our results confirm earlier findings that the time-lag between a perturbation and the resultant inner accretion rate variation depends on the frequency (or time-period) of the perturbation. Here we have quantified that the time-lag $t_{lag} \propto f^{-0.54}$, for time-periods less  than the viscous time-scale of the perturbation radius and is nearly constant otherwise. This, coupled with radiative process would produce the observed frequency dependent time-lag between different energy bands. We also confirm that if there are random Gaussian fluctuations of the $\alpha$-parameter at different radii, the resultant inner accretion rate has a power spectrum which is a power-law.
\end{abstract}


\end{frontmatter}



\section{Introduction}
Black hole X-ray binaries are powered by an accretion disc. The X-rays are emitted from the inner regions of the disc close to the black hole where the characteristic time-scales are expected to be of the order of less than seconds. Thus, it was rather surprising that the X-ray emission from these systems vary over  a wide range of time-scales. It is now believed that this wide range occurs because viscous fluctuations with long characteristic time-scales occur in the outer parts of the disc and subsequently these variations propagate to the inner parts \citep{5}. Such a model qualitatively explains the power-law form of the power spectra of the X-ray emission. Moreover, these long time-scale fluctuations arising from  large radii on propagating inwards  gets superimposed on the smaller time-scale fluctuations arising there. Hence, this explains the linear positive relation between the $r.m.s$ variability and the X-ray flux, that are observed in many X-ray binaries \citep{25,35,36,37,38,23,40}. Thus, the model provides a natural connection between the power spectra and the rms-flux relation which has been studied by several authors \citep[e.g.][]{24,40}. Moreover, the same process should be acting on accreting systems such as Active Galactic Nuclei and hence in general can be used to obtain the size scale of the system \citep{kel09,kel11}. By modeling the observed fluctuations, it may also be possible to obtain direct constraints on the hydrodynamic structure of the accretion disc \citep{tit07}.

Another critical aspect of the temporal behaviour of black hole X-ray binaries is that typically, the systems show hard lags such that the high energy photons are delayed relative to the soft ones. The time-lags  increase logarithmically with energy and are surprisingly frequency dependent, $t_{lag} \propto f^{-0.7}$ \citep{miy88,6,43}. The magnitude of these lags are too large for them to be due to light crossing time-scales or due to Comptonization unless the corona is unphysically larger than $10^{9}$ cms \citep{42,48}. Instead, the propagation model provides an interpretation where if softer photons are produced at a larger radii compared to the high energy ones, there will be a time-lag between them corresponding to the time take for the fluctuation to propagate from the outer radius to the inner regions \citep{30,31}. The energy and frequency dependence of these time-lags can also be explained within such a framework \citep{30,31}. For many years, it has been puzzling why high accretion rate systems do not seem to be affected strongly by the radiation pressure instability \citep{7,28}. The instability should have made bright black hole X-ray binaries in their soft state (when they are supposed to have a standard accretion disc) to have  X-ray flux variations of several orders of magnitude over time-scales of minutes. Interestingly, the absence of such variation could be due viscous fluctuations in the disc \citep{4}.

Given the success and the potential of such a propagation model to explain many of the timing properties of black hole X-ray binaries, there have been attempts to integrate the basic idea of the model with one which can also explain the time averaged spectrum and other timing features such as Quasi-periodic Oscillations (QPOs). A popular model to explain the hard X-ray emission from these sources during the hard spectral state, is that the standard accretion disc is truncated and the inner region is a hot flow \citep{44,45}. For such an accretion disc geometry, there have been a series of works \citep{24,46,47,Rap16}, which incorporate the basic effects of stochastic propagation model. Such an integrated approach, can explain the time-averaged spectra and the temporal properties such as rms-flux and time-lags arising due to propagation. The model can also explain the $\sim 1$ Hz QPO observed in these sources as relativistic precession of the hot flow. There is no doubt that in the future, there will be more such attempts to provide a comprehensive picture using more complex accretion geometries with the propagation model being an integral part of such an approach.

Since such generic models are complex it is natural that these initial endeavours that attempt to quantitatively explain the temporal features, include stochastic propagation in an approximate manner where it is assumed that the accretion rate varies at different radii and that the time taken for the propagation to move inwards is the local viscous time-scale. On the other hand, there are approximate analytical techniques which have been used to describe the diffusion process in the accretion flow and quantify the basic prediction of the propagating fluctuations \citep{8,5,30,31}. For example, while \citet{30} describes the propagation in terms of cylindrical sound waves, \citet{31} have used the approximate Green's function formalism of \citet{8} and \citet{5}. Solving the disc equations in a linearized form \citet{psa00} have shown that any transition radius in the disc will act like a low-pass filter to such fluctuations and have argued that resonant features may cause Quasi-periodic behaviour. It is possible, however, to solve the time dependent hydro-dynamical equations with a stochastic viscosity to obtain the full global non-linear behaviour of the disc. \citet{cow14} simulated such a disc with global stochastic fluctuations in the viscosity and confirmed  the non-linear behaviour of the $r.m.s$ being  proportional to the flux. They found that the accretion rate variations at frequencies lower than the local viscous frequency is coherent for different radii and hence revealing the basic idea of propagation fluctuation. They computed the expected time-lags and found that their magnitudes are smaller than the viscous time-scale and that they are frequency dependent. Further, \citet{hog16} performed global magneto-hydrodynamic simulation of accretion disc and confirmed that the Maxwell stresses in the low frequency regime, indeed produce fluctuations which propagate inwards. These simulation results, especially that the time-lags are a function of frequency need to be incorporated into future models which would strive to quantitatively explain the temporal behaviour of accreting systems.

In these previous works, the fluctuations have been introduced in the viscosity as inspired by magneto-hydrodynamic simulations \citep{hog16}. However, in the  standard $\alpha$-disc model \citep{3}, the viscosity is parametrised by $\alpha$ such that the viscous stress is $\alpha P$, where $P$ is the pressure. Viscous variations can be then represented by variation in the $\alpha$ parameter. Such simulations have been undertaken for, e.g., to study the effect of these variations on unstable radiation pressure dominated discs \citep{4} and to understand the effect of X-ray irradiation on the time dependent properties of the outer regions of an accretion disc \citep{27}. Since the nature especially the time-dependent behaviour of the turbulent viscosity is largely unknown, it is prudent to test different mechanisms by which the viscosity may fluctuate.

In this work we study the response of a  gas pressure dominated standard accretion disc to viscus fluctuations at different radii and time-periods, concentrating on the temporal behaviour of the accretion rate at different radii rather than on the emergent spectrum. One of our motivation is to confirm whether the salient features, especially the frequency dependent time-lag are similar when the fluctuations are introduced in $\alpha$ rather than the viscous stress. More importantly, the aim is to introduce simple empirical functions which capture the radial and frequency dependence of the
variability and time-lags which may then be used in models that quantitatively explain temporal features. The results obtained may perhaps be fairly generic despite the simplifying assumption of a standard gas pressure dominated disc. The expectation here is that empirical scaling laws obtained, may form the basis for more complex studies involving more complex accretion geometry and the coupling of the flow to the local radiative process.

In the next section, we list the basic time-dependent equations that describe the standard gas pressure dominated accretion disc. In \S 3,we study the response of such a  disc to sinusoidal perturbations at different radii and with different time-periods. In \S 5, we introduce Gaussian stochastic perturbations at each radii and study the temporal behaviour of the inner accretion rate. In the last section, we summarise and  discusses the important results of the work.

\section{Time-dependent disc structure equations} 

Since we are interested in the long term temporal evolution of an
accretion disc on viscous time-scales we assume as in the standard
disc, vertical hydro-dynamical equilibrium leads to 
\begin{equation}
P = \frac{GM\Sigma}{r^{3}}\frac{h}{2}
\end{equation}
where $h$ is the half thickness of the disc, $\Sigma = 2\rho h $ is
the surface density and the pressure $P$ is assumed to be due to gas
pressure only i.e. $P = {\Sigma} kT/{2\mu h m_{H}}$. Here $T$ is the
mid-plane temperature, $m_{H}$ is the mass of hydrogen atom and $\mu
\sim 0.5$ is the mean molecular weight.  The disc is also assumed to
be in thermal equilibrium such that the local viscous heat dissipated
is radiated. The viscous dissipation rate per unit area
 is taken in the standard form to be
\begin{equation}
Q_{vis}^{+} =\frac{3{\dot{M}}}{8\pi r^{2}}\frac{GM}{r}\left(1
-{\left(\frac{R}{r}\right)}^{0.5}\right)
\end{equation}
where $M$ is the mass of the black hole,
$R$ is the inner radius of
the disc, $\dot M$ is the accretion rate and $r$ is the radius of
disc. The disc is assumed to be radiating as a black body, hence the energy flux per unit area is given by \citep[e.g.][]{Sha86}
\begin{equation}
F(r)  \simeq \frac{\sigma T^{4}}{4\tau} = Q_{vis}^{+}
\end{equation}
where $T$ is the mid-plane temperature of the disc and $\sigma$ is the Stefan-Boltzmann constant.
The optical depth, $\tau$ is considered both due to electron-scattering, ${\bar{\kappa}_{es}}$ as well as due to free-free interaction, ${\bar{\kappa}_{ff}}$ and is given as

\begin{equation}
\tau = \Sigma({\bar{\kappa}_{ff}} + {\bar{\kappa}_{es}})
\end{equation}

Opacity due to free-free interaction, ${\bar{\kappa}_{ff}}$ is given by Kramer's opacity law written as
\begin{equation}
\bar{\kappa}_{ff} = 0.64\times {10^{23}}\rho T^{-7/2} cm^{2} g^{-1} \nonumber
\end{equation}
and the opacity due to electron scattering, 
${\bar{\kappa}_{es}} =  0.41$ cm$^{2}$ g$^{-1}$.

In the time dependent case, the surface density, $\Sigma$ changes with time if the accretion-rate, $\dot{M}$ varies with radius and in such a situation the disc behaviour is governed by the continuity  equation;
\begin{equation}
\frac{\partial{\Sigma}}{\partial t} + {\frac{1}{2\pi r}}{\frac{\partial{\dot{M}}}{\partial {r}}} = 0    
\label{sig}
\end{equation}
and the angular-momentum conservation equation needs to be retained in its differential form, 
\begin{equation}
\frac{\partial}{\partial t}\left({\Sigma 2\pi r}\sqrt{GMr}\right) + \frac{\partial}{\partial r}\left(\dot{M}\sqrt{GMr}\right)=\frac{\partial}{\partial r}\left(\alpha P 4\pi r^{2} h\right) \nonumber
\end{equation}
which can be simplified to
\begin{equation}
\dot{M} \frac{\partial}{\partial r}\left(\sqrt{GMr}\right)=\frac{\partial}{\partial r}\left(\alpha P 4\pi r^{2} h\right)
\label{mdot}
\end{equation}

The partial differential Equation (\ref{sig}) is solved numerically to
obtain the time dependent behaviour of an accretion disc using the standard
finite differencing scheme with the courant condition imposed as described
in \cite{27}.
At an outer
boundary radius, the disc values are fixed to the steady state ones
while for the inner one we maintain the viscous stress free
condition.

We find that after a transient period  the disc settles to a steady
state with a constant accretion rate $\dot{M_0}$ and having the steady state
surface density profile $\Sigma_0(r)$. A pertinent time-scale for the system 
at a given radius is
the viscous time-scale which is given by
\begin{equation}
T_{vis} = \frac{2\pi}{\dot{M_0}}\Sigma_0 {r^{2}}
\label{vistime}
\end{equation}
After this transient period which is roughly equal to few times the
viscous time-scale of the outer radius,  time varying
perturbations are introduced to the viscous parameter $\alpha$ as
described in the following sections, and the time dependent properties
of the disc are studied. For all the results presented in this work, we adopt $\alpha = 0.1$ and the
steady state value of the accretion rate to be 
$\dot{M_0} = 10^{17}$ g s$^{-1}$. 

\section{Sinusoidal perturbation at different radii}

\begin{figure}
\centering 
\includegraphics[width=0.77\textwidth,angle=0]{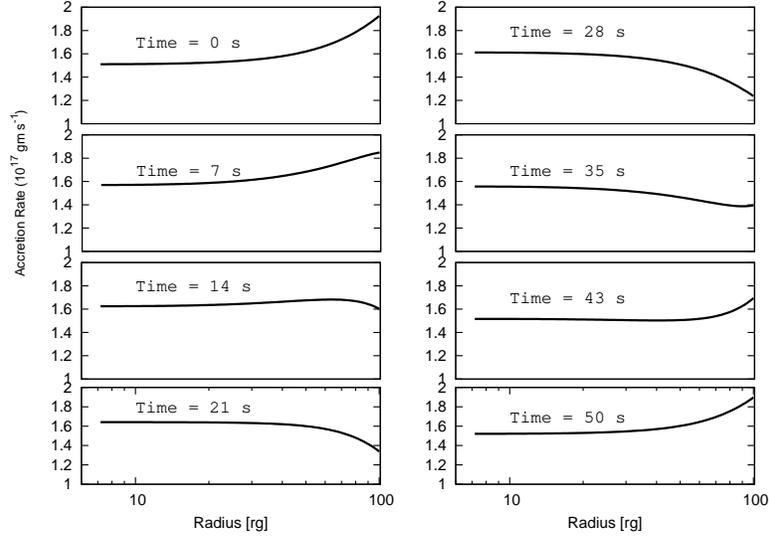}
\caption{Snapshots of accretion rate versus radii at different times. The perturbation has been introduced at the radius of 100$r_{g}$ with a time-period of 50 s.}
\label{movie}
\end{figure}

\begin{figure}
\centering
\includegraphics[width=0.55\textwidth,angle=270]{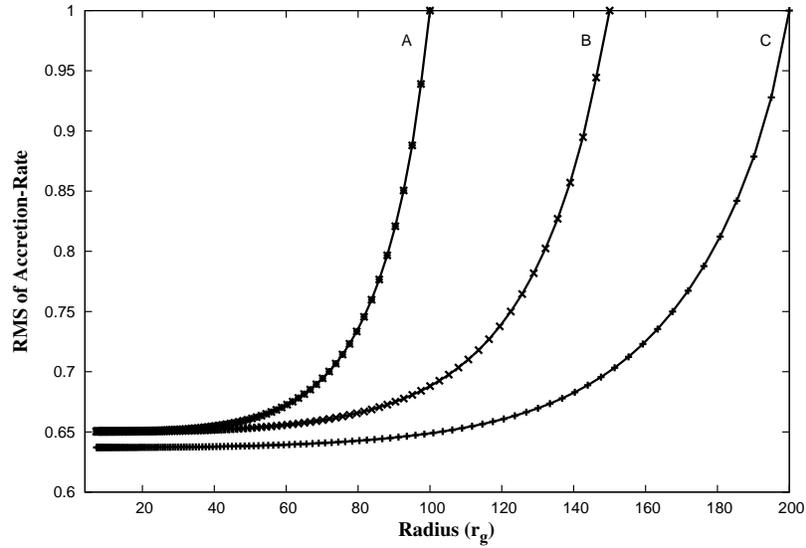}
\caption{The root mean square of the accretion rate variability as a function of radius. The $r.m.s$ have been normalised to that of the accretion
rate at the perturbation radius. The time-periods for the curves marked A, B and C are 0.2 times the viscous time-scale of the perturbation radius of 100, 150 and 200 $r_{g}$, respectively. The amplitude of the perturbation in all these cases decreases initially, but then saturates with radius.}
\label{rms_radius}
\end{figure}

We first study the response of the disc to a sinusoidal perturbation
of the viscous parameter $\alpha$ 
\begin{equation}
{\alpha(r_p,t)} = \alpha + \triangle{\alpha}sin \left(\frac{2\pi
  t}{T_{p}}\right)
\end{equation} 
at a particular perturbation radius, $r_p$ and with a time-period
$T_p$. The $\alpha$ variation causes the accretion rate to vary at
$r_p$ and the perturbation moves inwards to smaller radii. This is
illustrated in Figure \ref{movie} which shows snapshots in time for a
perturbation at $r_p = 100 r_g$ with $\triangle{\alpha}=0.01$. 
While the amplitude of the accretion
rate variation is large at $r_p$, it decreases as the perturbation
travels inwards. This is further illustrated in Figure
\ref{rms_radius}, where the root mean square ($r.m.s$) of the
accretion rate variability is plotted as a function of radii. Here the
time-period for the perturbation is 0.2 times the viscous time-scale
at the perturbation radius and the Figure shows the behaviour for
three different perturbation radii. The variability decreases rapidly
and then saturates at smaller radii. Such a behaviour is expected as
the perturbation gets damped when its time-period is smaller than the
local viscous time-scale and is unabated when the time-period is
larger. 

\begin{figure}
\centering

\includegraphics[width=0.55\textwidth,angle=270]{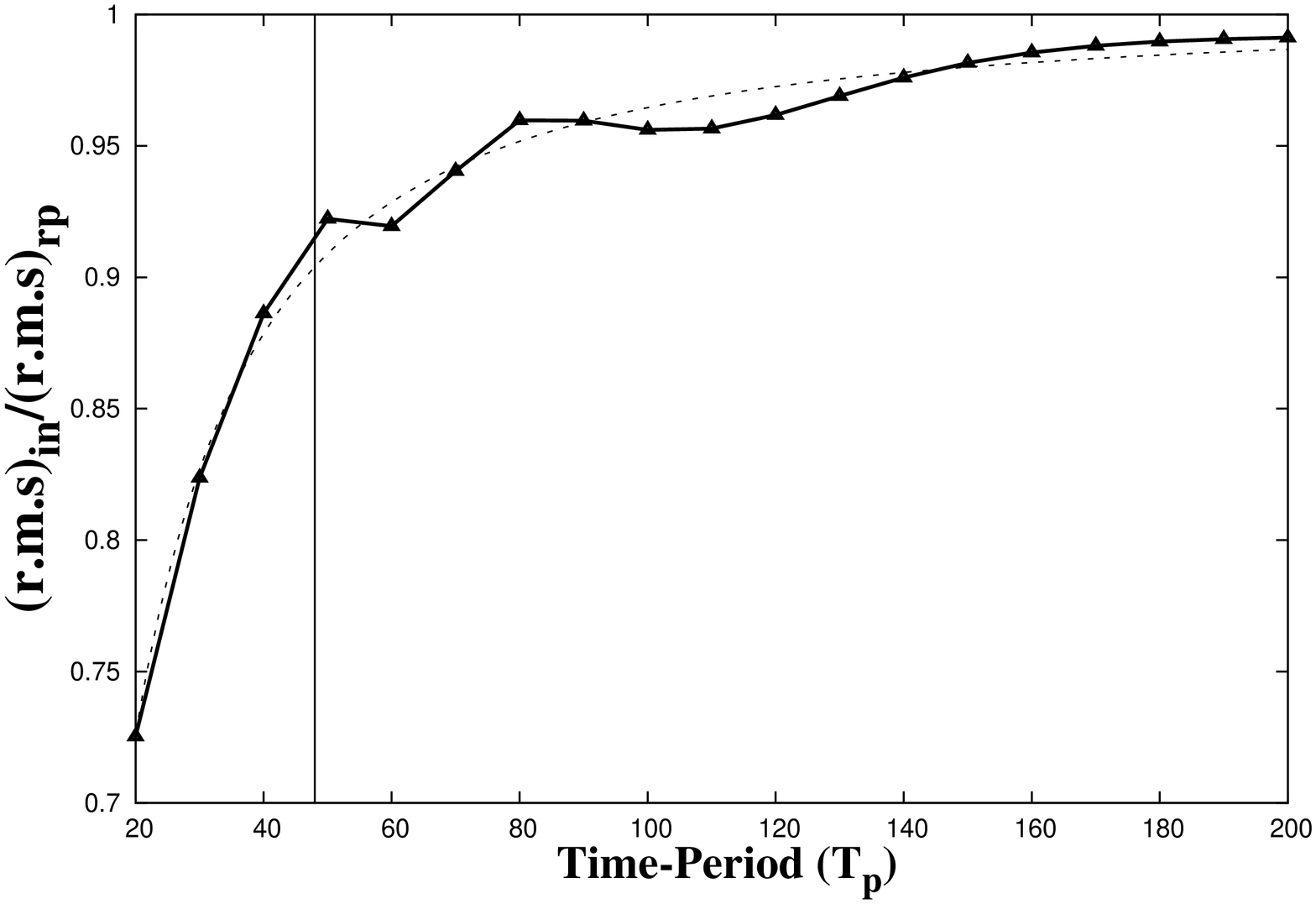}
\includegraphics[width=0.55\textwidth,angle=270]{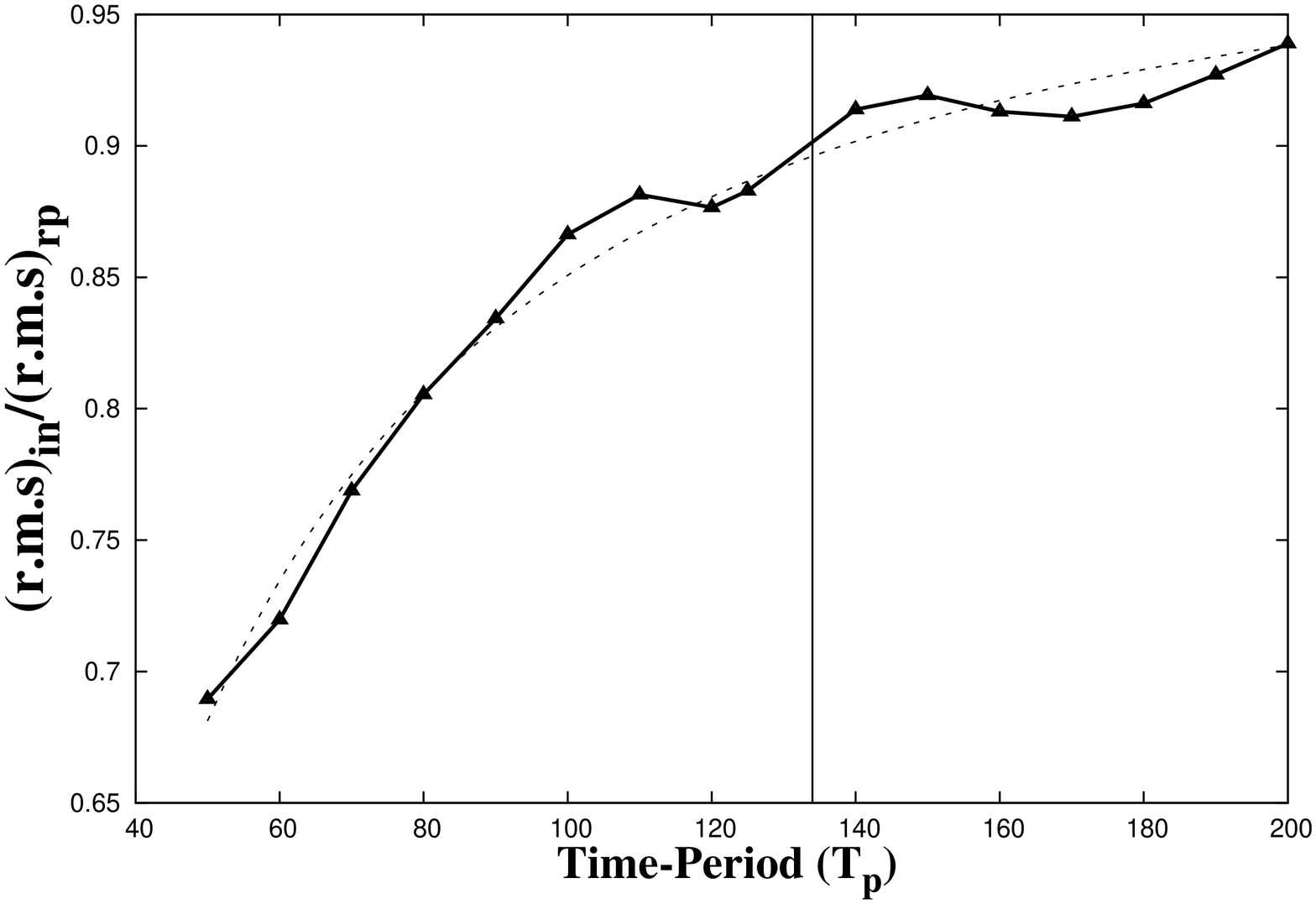}
\caption{The figures show the variation of ratio of the $r.m.s$ of accretion rate at the inner most radius (6$r_{g}$) normalised to the $r.m.s$ of accretion rate at the perturbation radius vs time-period, $T_{p}$. The top panel is for the case, when the perturbation radius, $r_{p}$ is at 100$r_{g}$, where the viscous time-scale, $T_{vis}$ is 48 seconds, which is marked by a vertical line. The lower panel is for the case, when the perturbation radius is at 200$r_{g}$, where the viscous time-scale is 134 seconds. Both curves have been fitted using a function of the form $\frac{1}{1+\left({AT_{vis}}/{T_{p}}\right)^{p}}$ (shown by dotted lines). The values of $A$ and $p$ are 0.21 (0.22) and 1.45 (1.42), respectively for 100$r_{g}$ (200$r_{g}$).}
\label{rms_ratio}
\end{figure}

The saturation of the $r.m.s$ at small radii, allows us to quantify the effect a perturbation has, by computing the ratio of the saturated $r.m.s$ (computed at the inner edge) with that at the perturbation radius $R = (r.m.s)_{in}/(r.m.s)_{r_p}$. This is illustrated in Figure \ref{rms_ratio}, where the ratio is plotted as a function of the time-period, $T_p$, when the perturbation radius are $r_p = 100$ (Top Panel) and $200$ $r_g$ (Bottom panel). The corresponding viscous time-scales for these radii are $\sim 48$ and  $\sim 134$ seconds, respectively and they are marked as vertical lines in the Figure. The amplitude ratio remains nearly constant at unity when the oscillation time-period is larger than the viscous time-scales and
decreases for smaller values. A phenomenological fit to the dependence of the ratio to the time-period can be 
\begin{equation}
R (T_p) = \frac{(r.m.s)_{in}}{(r.m.s)_{r_p}} = \frac{1}{1 + \left(\frac{A T_{vis}}{T_p}\right)^p}
\end{equation}

where $T_{vis}$ is the viscous time-scale at the perturbation radius, $r_p$. Comparison  with the numerical values show that $A =$  0.21 and 0.22 and $p =$ 1.45 and 1.42 for $r_p = 100$ and $200$ $r_g$ respectively, and are represented by dotted lines in Figure  \ref{rms_ratio}. Thus, this functional form captures the dependence of the inner disc variation to an outer disc fluctuation. 

\section{Time-lag for sinusoidal perturbations}

It is expected that a perturbation at a radius $r_p$ would propagate inwards and reach the inner radii after a time-lag equal to the viscous time-scale at $r_p$. There are a couple of reasons this expectation might not be met. First, the viscous time-scale computed using equation (\ref{vistime}) is after all an approximation and hence the actual time-lag associated with the perturbation may differ by a factor of few. More importantly, in a cylindrical geometry, one can expect that the time-lag should have a dependence on the time-period (or frequency) of the perturbation. Unlike
planar waves, the speed of cylindrical sound waves depends on its frequency, which has been exploited to explain frequency dependent time-lags in Cygnus X-1 \citep{30}. The diffusion equation that describe the propagation here should also have this property and we test whether this is the case, by computing the time-lag between the perturbation and its response in the inner disc region.

\begin{figure}
\centering
\includegraphics[width=0.50\textwidth,angle=270]{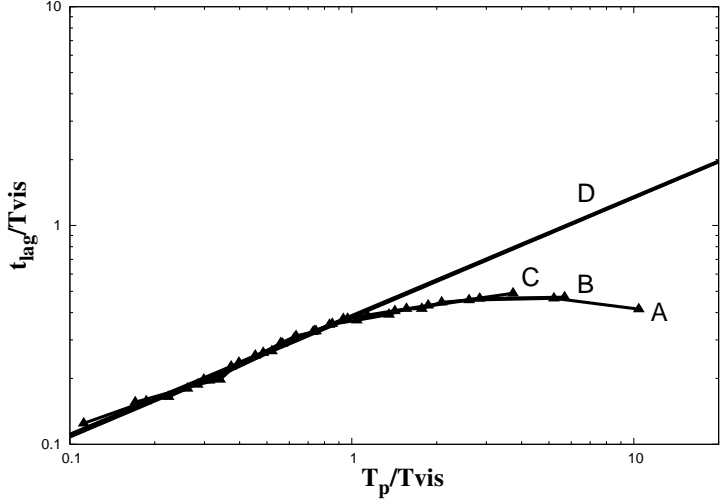}
\caption{The figure shows the variation of the time-lag, $t_{lag}$ with time-period, $T_{p}$, both normalised to the viscous time-scale, $T_{vis}$, corresponding to the radius where sinusoidal perturbation of different time-periods are given. The curves marked A, B and C correspond to the perturbation radii of 100, 150 and 200 $r_{g}$, respectively. All these curves are represented by $\frac{t_{lag}}{T_{vis}} \sim 0.38 \left(\frac{T_{p}}{T_{vis}}\right)^{0.54}$ shown in the figure by curve D. This function was fitted only for $\frac{T_{p}}{T_{vis}} < 1 $.}
\label{timelag}
\end{figure}

Figure \ref{timelag} shows the computed time-lag, $t_{lag}$ divided by the viscous time-scale, $T_{vis}$ of the perturbation radius versus the time-period, $T_{p}$ of the oscillation which is also divided by the same viscous time-scale. The curves marked A, B and C correspond to the perturbation radii 100, 150 and 200 $r_g$ respectively. In this representation, all three curves line up along each other and show that in general, for time-periods longer than the viscous time-scale, the time-lags are nearly independent of the time-period and are roughly 0.4 times the viscous time-scale. For shorter time-periods the time-lags decrease proportionally with the  time-period. A functional form for the time-lag for time-periods less than viscous time-scale is 
\begin{equation}
\frac{t_{lag}}{T_{vis}} \sim 0.38 \left(\frac{T_{p}}{T_{vis}}\right)^{0.54}
\end{equation}
We note this result that the time-lag increases with the time-period is consistent with those found by \citet{cow14}.

\section{Multi-Gaussian perturbation at different radii}

In an accretion disc, it is expected that there will be stochastic perturbations at all radii which  propagate inwards, causing the inner accretion rate to vary in a wide range of frequencies. To get insight into the nature of such a system we conduct the following numerical toy experiment.

We consider a disc with an outer radius of 35 $r_g$ where the
corresponding viscous time-scale is 10 seconds. We divide the disc in
64 bins which are logarithmically spaced in radii. At each radial bin,
the viscous parameter $\alpha$ is perturbed by $\Delta \alpha$ where
$\Delta \alpha$ is derived from a Gaussian distribution having 
a width of 0.0005. A narrow distribution has been chosen 
because for a larger width, there are occasionally large
$\alpha$ variations. For these large variations, the numerical code requires
very small time steps and hence leads to very time consuming computations.
The perturbation is given after a time equal to the local viscous
time-scale. Thus, $\alpha$ at each radial bin, is forced to vary in a
step like fashion where the change occurs on a time equal to the local
viscous time-scale. This step like variation in $\alpha$ causes a
spike in the local accretion rate whenever $\alpha$ is suddenly
changed. 

\begin{figure}
\centering
\includegraphics[width=0.425\textwidth,angle=270]{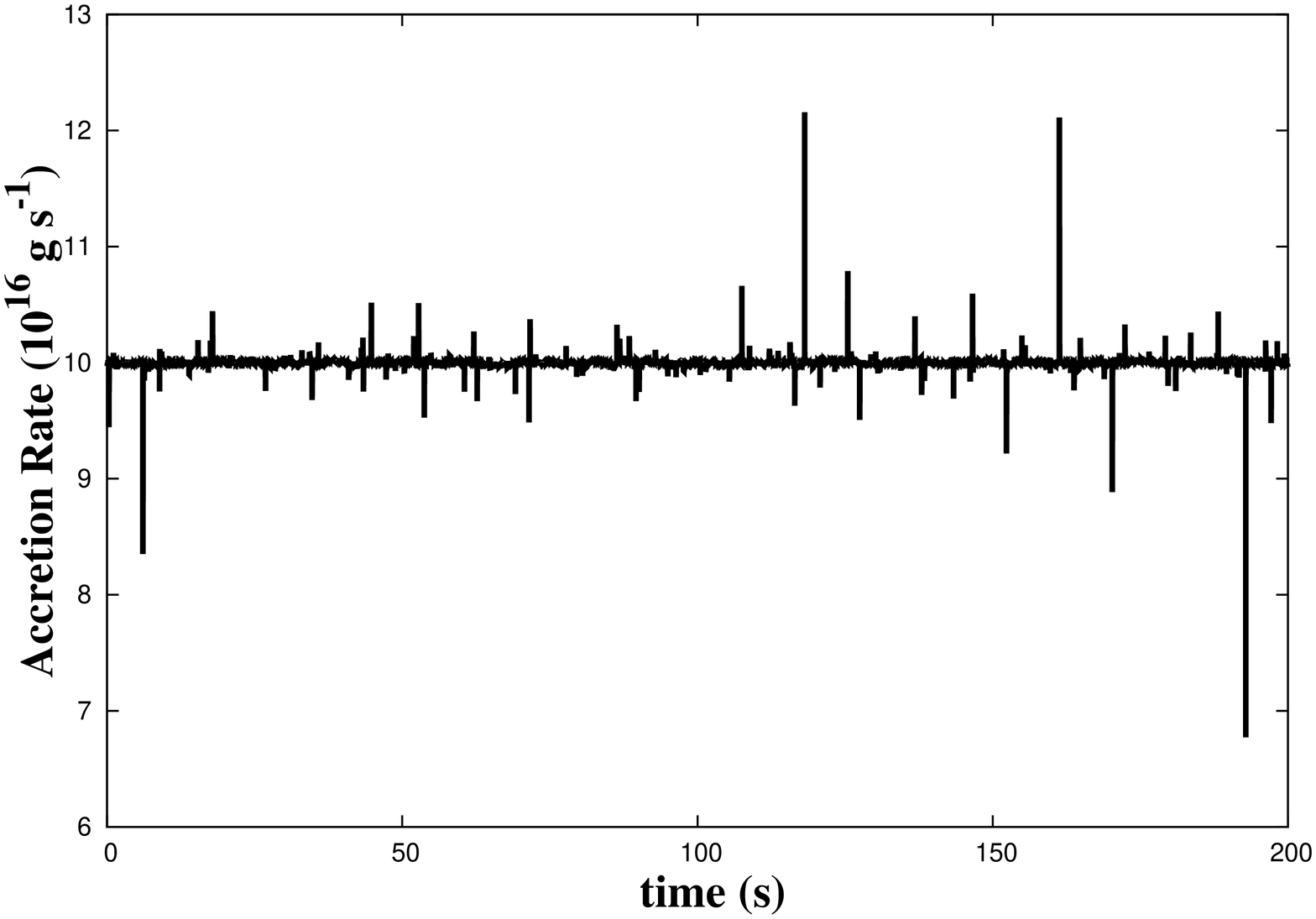}
\includegraphics[width=0.425\textwidth,angle=270]{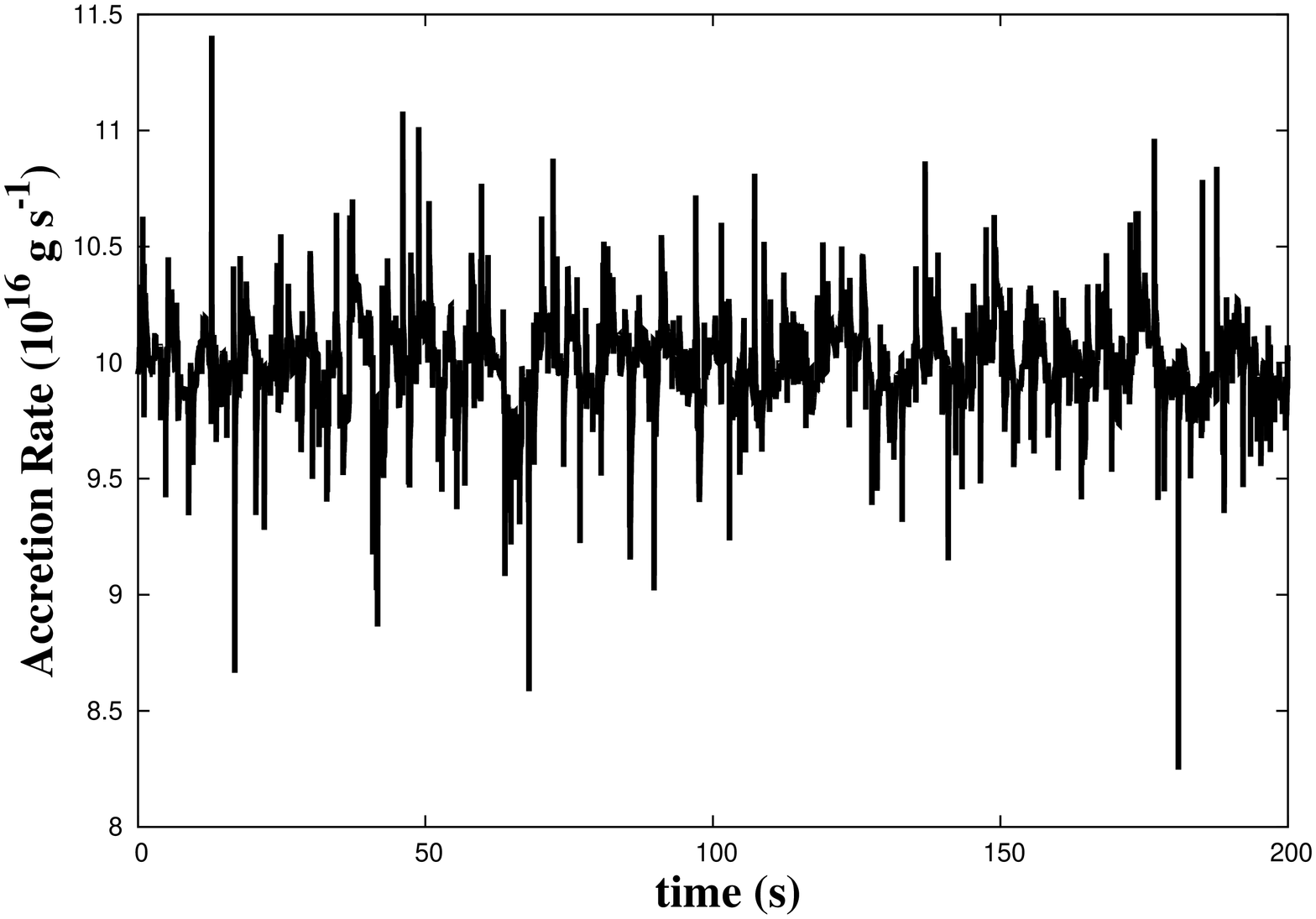}
\includegraphics[width=0.425\textwidth,angle=270]{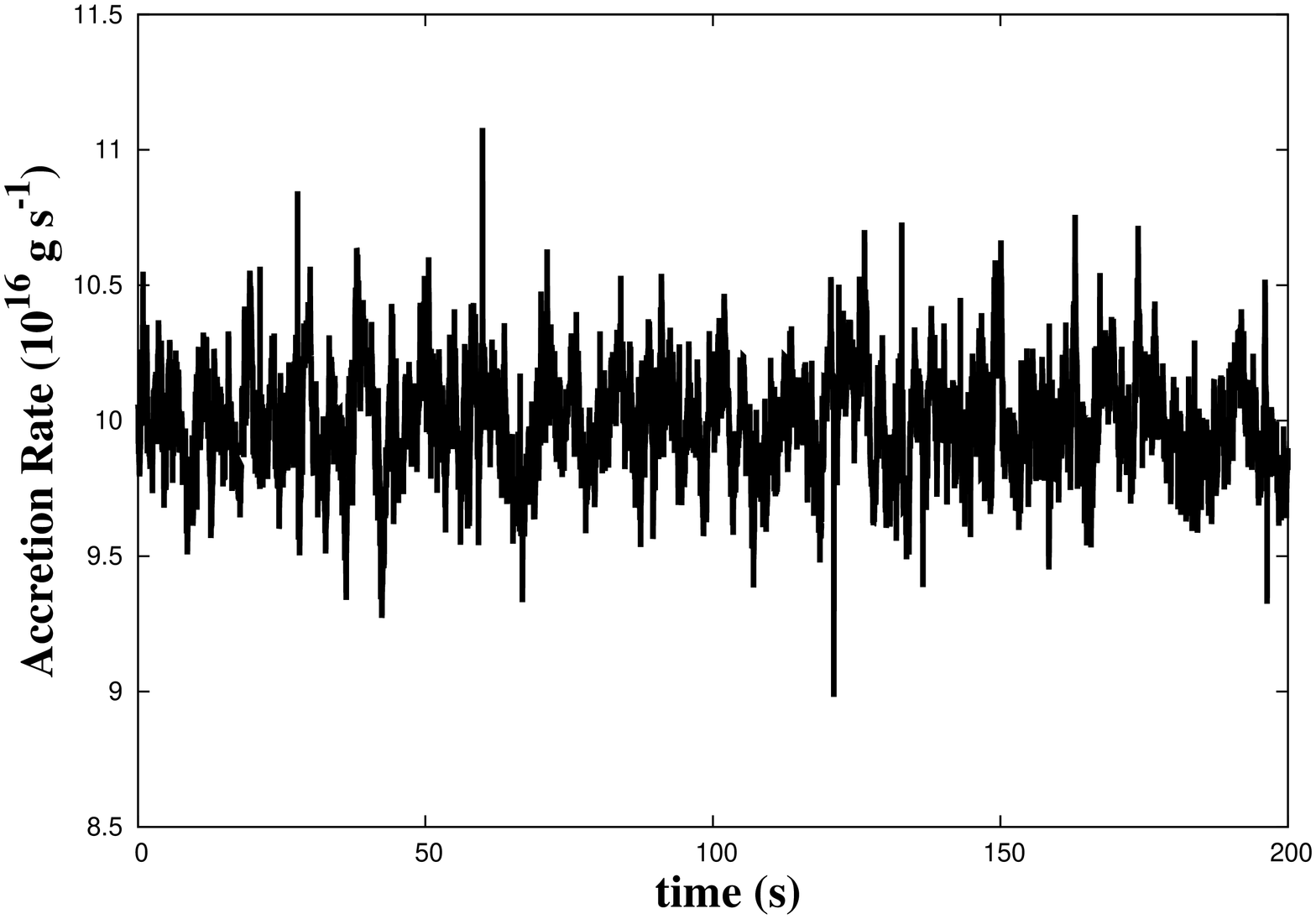}
\caption{The figure shows the variation of accretion-rate, $\dot{M}$, when random Gaussian perturbations are added to the viscosity parameter, $\alpha$ at different radial points of the disc. The curves on the top, middle and lower panel show the variation of accretion rates, $\dot {M}$ with time near the outer radius ($34r_{g}$), in between radius ($21r_{g}$) and at the inner radius ($7r_{g}$), respectively.}
\label{accret}
\end{figure}

Near the outer radii (i.e. at two radial bins away from the last one), 
this causes the magnitude of the change
in the accretion rate to have a sharp rise and fall as shown in the
top panel of Figure \ref{accret}. The accretion rate variations at this radius,
 correspond to
the local $\alpha$ variation at that bin as well as the variation 
induced in the outer two bins. The time-scales for these spikes are
approximately ten seconds i.e. the
local viscous time-scale at 35 $r_g$. 

\begin{figure}
\centering
\includegraphics[width=0.50\textwidth,angle=270]{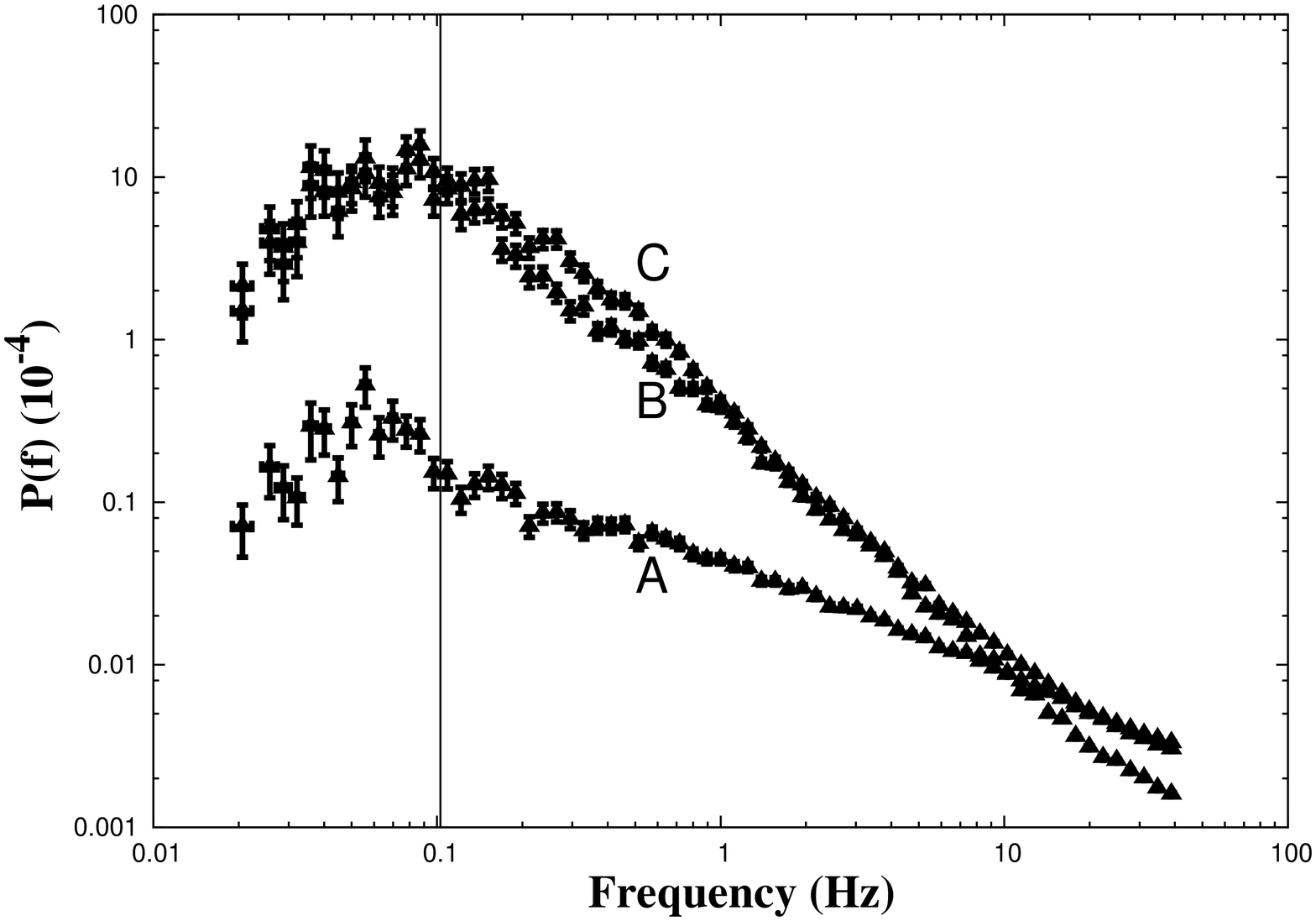}
\caption{The curves A, B and C shows the power spectrum corresponding to the radii, whose accretion-rate variations with time are shown in the upper, middle and lower panels, respectively in Fig 5. The spectral index for curve A is $\sim 0.6$, while for curves B and C, is of the order of $\sim 1.5$. The vertical line is the inverse of the viscous time-scale corresponding to $35r_{g}$.}
\label{pow}
\end{figure}

For smaller radii, the accretion rate varies due to the local step-like variation of $\alpha$ and due to the propagation of the fluctuations from the outer regions. The crucial element here is that the two causes for the local accretion rate variation are multiplicative rather than being additive. Thus a significant enhancement of the variability occurs at smaller radii as compared to the larger ones. The accretion rate variation in an intermediate (21$r_g$) and at an inner radius (7$r_g$) are shown in the middle and bottom panels of Figure \ref{accret}. As expected the variability increases as the radii decreases. This is quantified further in Figure \ref{pow} where the power spectra of the accretion rate variability of the three radii considered in Figure \ref{accret} are shown and marked as A, B, and C. While the power spectra of the  outer radius (Curve A at 34$r_g$) is weaker, there is enhanced power in the inner radii (Curve B and C at 21 $r_g$ and 7$r_g$). The inner radii power spectra have a power-law like behaviour beyond 0.1 Hz with an index of $\sim 1.5$. We have verified that changing the number of annular bins does not cause any qualitative change in the power spectra.

We caution against over interpretation of the results of this  numerical toy experiment, since for computational purposes the region of disc is rather small and the perturbations are given in a rather ad-hoc manner. Nevertheless, some of the essential elements of the propagation fluctuation model seem to be captured, i.e. the power spectra of the accretion rate in the inner regions has a power-law form and since the effect is multiplicative, the variability is stronger in the inner region than in the outer.

\section{Summary and discussion}

We solve the time-dependent equations for a standard gas pressure dominated accretion disc with perturbations on the local viscous parameter $\alpha$ to understand the response of the accretion rate at different locations of the disc.

For a sinusoidal perturbation at an outer radius $r_p$, we find that the ratio of the amplitude of the accretion rate variation at the inner disc to that at the perturbation radius, can be empirically represented by
\begin{equation}
R (T_p) = \frac{(r.m.s)_{in}}{(r.m.s)_{r_p}} = \frac{1}{1 + \left(\frac{0.2 T_{vis}}{T_p}\right)^{1.4}} \nonumber
\end{equation}
where $T_p$ is the time-period of the sinusoidal perturbation and $T_{vis}$ is the viscous time-scale at the perturbation radius. We find that the time-lag between the inner accretion rate variability with respect to that of the perturbation radius, depends on the time-period of the oscillation and can be empirically represented as
\begin{equation}
\frac{t_{lag}}{T_{vis}} \sim 0.38 \left(\frac{T_{p}}{T_{vis}}\right)^{0.54} \nonumber
\end{equation}
for $T_P < T_{vis}$ and is nearly a constant for larger values.

We further show by having all along the disc, step-like fluctuations of the viscous parameter $\alpha$ on the local viscous time-scale, that the power spectrum of resultant inner accretion rate variability has a power-law shape in frequency with index $\sim 1.5$. 

The results obtained are consistent with those found by \citet{cow14} which show that the qualitative response of the disc is same when the fluctuations are introduced in the $\alpha$ or directly to the viscosity which implies that perhaps they do not depend on the details of the viscosity prescription. The empirical functions obtained here, especially the dependence of the time-lag on frequency, will have utility in making models that predict the power spectra and time-lags of X-ray binaries within the framework of propagating fluctuations such as those described in \cite{47,Rap16}. In particular, the dampening of the fluctuation can be expressed using the empirical relation above and the time-lag between the accretion rate variation can be taken to be time-period (or frequency) dependent. These will no doubt have an impact on both the qualitative and quantitative interpretations that such models make.

However, it should also be emphasized that the results here are based on simple assumptions which may not be valid in real accretion discs in X-ray binaries. In X-ray binaries, the observed variability is often seen and quantified in the X-ray emission arising from a corona on top of a disc or from a hot inner region where the standard disc is truncated. In both cases one needs to consider a more generalized disc than the standard gas pressure dominated one assumed here. A perhaps important omission is that the pressure support in the radial direction has been neglected, which while true for low luminosity discs, would be violated in high luminosity ones and if there is a hot inner disc. Radial pressure support will give rise to radially moving sound waves which may significantly alter the results obtained here.

In conclusion, while we verify by solving the time-dependent disc equations the salient features of the propagation fluctuation model, we also provide an empirical relation for the time-period dependent time-lag between the accretion rate fluctuations. Clearly, there is a scope to do more advanced computation using more detailed physics with predictive powers. Such analysis will provide insight into the geometry and nature of these systems, when compared with the existing  data and future ones expected from the newly launched {\it AstroSat}.

\section{Acknowledgments}
The authors are grateful to ISRO-RESPOND Program for providing the financial assistance under grant No. ISRO/RES/2/370 for carrying out this work. Four of us (NA, NI, BM and MH) are highly thankful to IUCAA, Pune for providing the computing and library facilities in completing this manuscript.\par





\end{document}